# Observations contradict galaxy size and surface brightness predictions that are based on the expanding universe hypothesis

Eric J. Lerner[★]

*LPPFusion, Inc., 128 Lincoln Blvd., Middlesex, NJ 08846-1022, USA*

**ABSTRACT**
In a non-expanding universe, surface brightness is independent of distance or redshift, while in an expanding universe it decreases rapidly with both. Similarly, for objects of the same luminosity, the angular radius of an object in a non-expanding universe declines with redshift, while in an expanding universe this radius increases for redshifts $z > 1.25$. The author and colleagues have previously shown that data for the surface brightness of disc galaxies are compatible with a static universe with redshift linearly proportional to distance at all $z$ [static Euclidian universe (SEU) hypothesis]. In this paper, we examine the more conventional hypothesis that the universe is expanding, but that the actual radii of galaxies of a given luminosity increase with time (decrease with $z$), as others have proposed. We show that the radii data for both disc and elliptical galaxies are incompatible with any of the published size-evolution predictions based on an expanding universe. We find that all the physical mechanisms proposed for size evolution, such as galaxy mergers, lead to predictions that are in quantitative contradiction with either the radius data or other data sets, such as the observed rate of galaxy mergers. In addition, we find that when the effect of telescope resolution is taken into account, the $r$–$z$ relationships for disc and elliptical galaxies are identical. Both are excellently fit by SEU predictions. An overall comparison of cosmological models requires examining all available data sets, but for this data set there is a clear contradiction of predictions based on an expanding universe hypothesis.

## 1 INTRODUCTION

As Tolman (Tolman 1930, 1934) demonstrated, in any expanding cosmology, the surface brightness (SB) of any given object is expected to decrease very rapidly with $z$, being proportional to $(1 + z)^{-3}$, where $z$ is the redshift and where SB is measured in AB magnitudes. By contrast, in a static (non-expanding) universe, where the redshift is due to some physical process other than expansion (e.g. light-aging), the SB is expected to be strictly constant when AB magnitudes are used. The Tolman test thus tests for the reality of universal expansion. Previously, the author and colleagues (Lerner 2006, 2009; Lerner, Falomo & Scarpa 2014) have demonstrated that extensive SB data for disc galaxies from *Galaxy Evolution Explorer* (*GALEX*) and *Hubble Ultra-Deep Field* (HUDF) is entirely compatible with a static universe where $z$ is linearly proportional to distance for all $z$. (A hypothesis of the relation of distance and $z$ in a non-expanding universe is necessary in order to convert apparent magnitudes into absolute magnitudes and thus to compare the SB of galaxies with the same absolute luminosity.) That is, the observed SB is independent of $z$. In this earlier work, we also show how other authors' data are compatible with this conclusion (Lerner et al. 2014).

A justification for using the linear hypothesis that $d = cz/H_0$ in testing the static model is presented in earlier work (Lerner et al. 2014). First, this relationship is well-confirmed in the most local universe. Secondly, the flux–luminosity relation derived from this assumption is remarkably similar numerically to the one found in the concordance cosmology, the distance modulus being virtually the same in both cosmologies for all relevant redshifts. Thus, the fit of the linear hypothesis to the supernovae Type Ia (SNe Ia) data is close to the fit for concordance cosmology, differing by at most 0.2 magnitudes for $z < 4.5$. For the range of $z$ covered by SNe Ia, the closeness of the fits has been improving with recent revisions in the concordance values. Since the linear hypothesis fits the SNe Ia and local universe distance-redshift data, it can be considered as a phenomenological hypothesis, and at this time no physical model is proposed to explain the linear relation.

The question we address in this paper is whether the same data can also be explained by an expanding universe hypothesis, with the additional hypothesis of evolution of the size of galaxies over time. Comparing the SB of galaxies of a given absolute luminosity is mathematically identical to comparing the radii of these galaxies. In a non-expanding universe, the prediction is simple – the SB is independent of $z$, and thus the radii of galaxies of a given luminosity are also constant with $z$. The angular radius in this hypothesis is linearly proportional to $1/z$. However, in an expanding universe, the same data can possibly be explained by a growth in radius of galaxies with time or, equivalently, a decrease in radius with $z$. The decrease in radius would have to occur in such a way as to cancel out the SB decrease, and the accompanying increase in observed angular diameter predicted by the expanding universe hypothesis. In this hypothesis, the accord between the data and the static universe hypothesis would then be considered to be a mere coincidence.

★ E-mail: eric@LPPFusion.com

A number of authors have proposed physical mechanisms for size evolution of galaxies (Mo, Mao & White 1998; Fan et al. 2008; Hopkins et al. 2009; Naab, Johansson & Ostriker 2009). We will compare these predictions with the data for both disc and elliptical galaxies. In this way we will conclude which predictions that have been made on the basis of the two cosmological hypotheses are compatible with the galaxy radius and SB data set.

## 2 SIZE EVOLUTION HYPOTHESIS FOR DISC GALAXIES

In previous work (Lerner et al. 2014), Lerner et al. examined data for high-luminosity disc galaxies as observed in rest frames near-ultraviolet (NUV) and far-ultraviolet (FUV) wavelengths. Such galaxies have young stellar populations and thus formed a relatively short time before the time at which they are observed. For the expanding universe hypothesis, Mo, Mao and White (Mo et al. 1998) first showed that the radius of disc galaxies forming at redshift $z$ should be a fixed fraction of the size of the dark matter halo. This, in turn, is proportional to $H^{-1}(z)$ for a fixed virial velocity or $H^{-2/3}(z)$ for a fixed mass, and somewhere in between for a fixed absolute luminosity $L$, where

$$H(z) = H_0[\Omega_m(1+z)^3 + \Omega_k(1+z)^2 + \Omega_\Lambda]^{1/2}, \quad (1)$$

where $\Omega_m$ is the ratio of matter density to closure density, $\Omega_\Lambda$ is the ratio of dark energy density to closure density, and $\Omega_k$ is the curvature parameter, it is assumed to be zero for an inflationary universe.

For a non-expanding universe, SB is independent of $z$, and thus for a given luminosity the radius of disc galaxies should be constant with $z$.

To test the concordance, expanding-universe hypothesis, we first compare the predictions with the published data from large catalogues of *Hubble Space Telescope* (*HST*) observations analyzed by Shibuya et al. (2015). This is the largest study to date of the sizes of disc galaxies. It is derived from a range of data sets, including HUDF09, GOOD-S and GOOD-N, Deep and Wide, UDS, AIEGIS COMOS, and HFF. For a low-$z$ comparison, we use our own previously published data from *GALEX* (Lerner et al. 2014). We compare the *GALEX* FUV data, observed at 155 nm, and the observations in Shibuya observed at 150 nm in the rest frame and labelled 'UV' in that source. In order to have sufficient number of low-$z$ galaxies, we select the UV absolute magnitude range centred at $MUV = -18$. We use in both cases the median radius of the samples for comparisons.

The $r_e$ reported by Shibuya et al. are 'circularized', meaning that the radii measured by the semi-major axis are divided by $q^{0.5}$, where $q = a/b$, $a$ being the major-axis and $b$ the minor-axis lengths.

**Table 1.** Median radii of bright disk galaxies (concordance assumptions).

| Z | Median $r$ circ (kpc) | $1/q^{0.5}$ | median $r$ (kpc) |
|---|---|---|---|
| 0.027 | | | 6.58 |
| 1.5 | 0.94 | 1.65 | 1.551 |
| 2.5 | 0.685 | 1.69 | 1.15765 |
| 3.5 | 0.572 | 1.61 | 0.92092 |
| 4 | 0.509 | 1.68 | 0.85512 |
| 5 | 0.506 | 1.61 | 0.81466 |
| 6 | 0.371 | 1.61 | 0.59731 |
| 8 | 0.356 | 1.61 | 0.57316 |

*Notes.* Column (1): mean redshift for each sample. Column (2): median circularized galaxy radius from Shibuya et al. (2015). Column (3): mean conversion factor, where $q = a/b$; $a$ is semi-major axis and $b$ is semi-minor axis. Column (4): median non-circularized galaxy radius obtained by multiplying column (2) by column (3). First row of table is from Lerner (2014).

However, the $r_e$ in the *GALEX* comparison sample are not circularized, but are based on the total light contained within a circular radius. This scales of course as the semi-major axis. In order to convert circularized into non-circularized radii, we divided the median non-circularized radii by the circularized radii for $z$-range subsets of the data from Shibuya et al. We then multiplied these ratios, which vary very little over the $z$-ranges observed, by the circularized radii, to obtain to non-circularized radii. The results are presented in Table 1.

It is useful to compare these results and those the author and collaborators earlier obtained with a smaller HUDF sample, using somewhat different techniques of measurement, but based on the same telescope. This will give an indication of the robustness of the results. The earlier Lerner et al. results were published as SB, but we here present the median angular radii from the same subsets, both those observed at rest wavelength bands centered at 155 nm (FUV) and those centered at 230 nm (NUV). The high-$z$ sample used, described in greater detail in the previous paper (Lerner et al. 2014), consists of disc galaxies from the HUDF images with $M$: $-17.5 < M < -19.0$ as observed in the FUV and NUV rest-frame bands.

We obtain angular radii from the published Shibuya et al. results by applying the standard angular distance conversion, using the concordance cosmology assumed by Shibuya et al. The comparison is shown in Fig. 1. As can be seen, the two analyses are in excellent agreement, so we can consider the results to be robust with respect to the exact algorithms for measurement of the radii.

We now can compare the results with the predictions of the Mo et al. theories. To do so, we need to add a sample at low redshift. We use the sample at $z = 0.027$ using *GALEX* galaxies observed in the FUV band from Lerner et al.

The results are presented in Fig. 2. We use only fits that go through the point at $z = 0.027$, so the error bars on each point include the error in this point. Clearly most of the points lie far from the theoretically predicted line. With a chi-square value of 121 relative to the steepest predicted slope of $-1$ of log $r$ on log $H(z)$, the probability of a fit to the Mo et al. theory is virtually zero. This theory is excluded by the data.

Even if we ignore the need for a physical mechanism for size evolution and simply test if the data can be fit by *any* evolution of the form $r = H(z)^{-n}$, where n is any constant, we still cannot obtain a fit to the data. The formal best fit for $n$ is $n = 1.13$ and the chi square of the data relative to this fit is 78, also essentially zero probability. This hypothesis is thus excluded at a $5\sigma$ level.

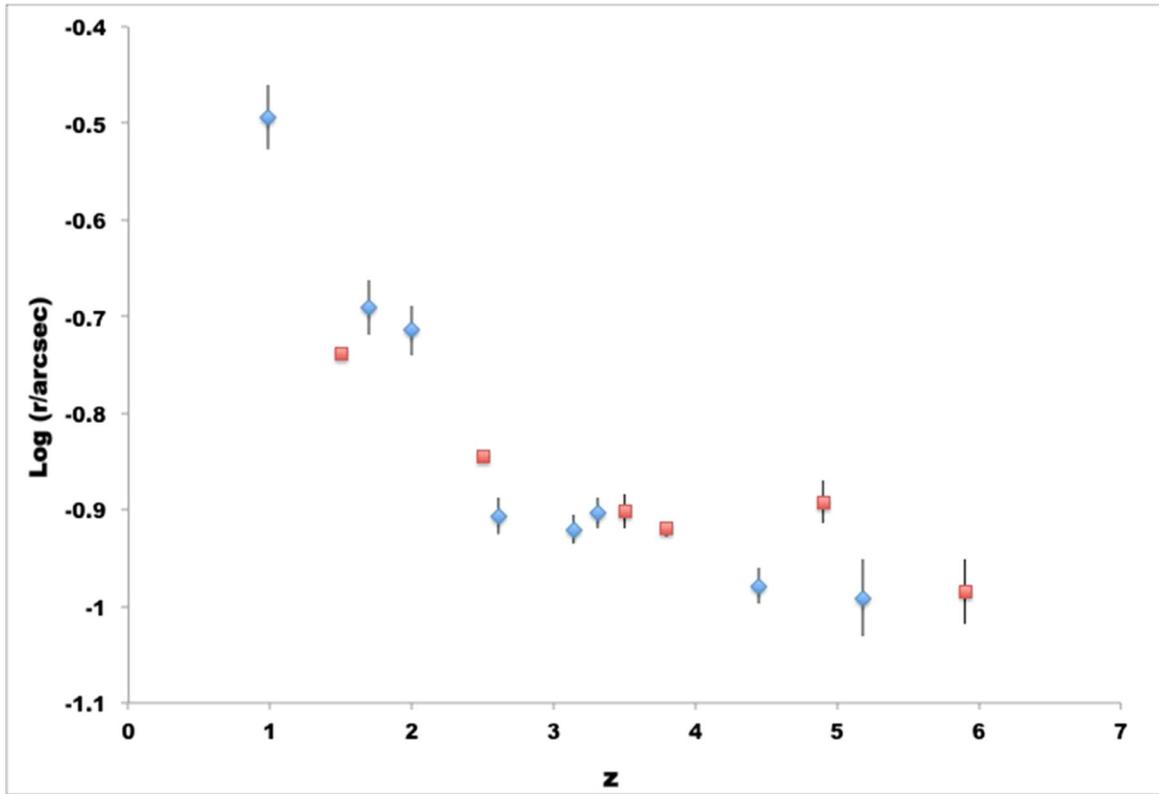

**Figure 1.** The log of the median angular radius of UV-bright disc galaxies samples with M centered at –18 from Shibuya et al. 2015 plotted against redshift (red squares) are compared with similar samples from Lerner, Scarpa & Falomo 2014 (blue lozenges). The two results are in close agreement.

It is important to understand why these comparisons do not arrive at the same conclusions as some other work comparing disc galaxy size evolution to the Mo et al. theory. For example, van der Wel et al. (van der Wel et al. 2014) find a good fit of disc galaxy size to $H(z)^{-0.66}$. Shibuya et al. find a good fit to $H(z)^{-0.97}$. Here, the data show no good fit of this form at all. Why are such different conclusions derived from what should be very similar data sets?

A key reason is the omission, in almost all these studies, of comparison galaxy sets at very low $z$. As can be seen from Fig. 2, the slope of log $R$ on log $H(z)$ steepens swiftly as $z = 0$ is approached. If the point at or near $z = 0$ is excluded, the slope of the curve is greatly reduced and it more closely approximates a straight line. Using the data presented in Fig. 2, without the point at $z = 0.027$, we obtain a slope of log $R$ on log $H(z) = -0.55$, close to and a bit below the range predicted by Mo et al. The chi-square measure declines to 2.78, an excellent fit.

However, omitting the point at low $z$ leads to erroneous conclusions. To test the various cosmological conclusions, we have to measure how size changes with increasing distance or look-back time *from the conditions at present*. That is what the predictions predict. To fit the data by excluding the nearest galaxies invalidates the comparisons. Rather, a fit has to be required to go through the low-$z$ point, within the statistical uncertainties of that point. It is when the fit is required to fit the point at or near the origin that $H(z)^{-n}$ fits become totally unacceptable. The $H(z)^{-0.55}$ curve fits $z = 0$ galaxies that are 2.7 times smaller than the actual galaxies.

Closely associated with this error is the common mathematical error in the literature of confusing Mo et al. predictions of $H(z)^{-n}$ with the mathematically very different fits of $(1 + z)^{-n}$. For example, Shibuya et al., among many others, states that $(1 + z)^{-1.5}$ and $(1 + z)^{-1}$ correspond to the case of fixed viral mass and fixed circular velocity. However, as can be seen in Fig. 2, mathematically the two functions $(1 + z)^{-(1.5n)}$ and $H(z)^{-n}$ give similar slopes predictions only for $z > 1$. For $z < 1$, the slopes diverge, so that for curves that go through the low-$z$ points – as any valid prediction must – the radii at high $z$ differ by a factor of $\Omega_M^{1/2}$ or 1.82. Thus, data sets that fit $(1 + z)^{-n}$ are unacceptably poor fits to $H(z)^{-n}$.

For the data presented here, even functions of the form $(1 + z)^{-n}$ are not good fits, although far better than the *Mo* et al. predictions. The best fit here is $n = 1.31$ and the chi square is 26.8, for a probability of only $6.2 \times 10^{-5}$, or $3.8\sigma$. In any case, no theory actually predicts functions of this form.

There are additional reasons for the wide range of fits to the disc galaxy data, connected with the resolution-size (RS) effect that is discussed in the next section. These additional reasons are detailed in Section 7.

## 3 RESOLUTION-SIZE (RS) EFFECT

If we want to compare the same observation of disc galaxy size with the predictions of the non-expanding static Euclidian universe (SEU) hypothesis, we must take into account the effects of telescope resolution. If a galaxy's angular size is close to or smaller than a telescope's resolution, its size will be measured as larger than when the same galaxy is observed with a telescope that has much better resolution, even when the PSF is modelled. We term this the 'RS effect'.

This effect is important in testing the SEU hypothesis, since this hypothesis predicts that angular radius decreases linearly with increasing $z$, so distant galaxies of the same linear size will have far smaller angular radii. Fortunately, there exist data sets that allow the linear resolution of low-$z$ galaxies observed with low-resolution telescope to be matched with the linear resolutions of high-$z$ galaxies observed with the *HST*. Such matching is model-dependent but, since the SEU makes unambiguous, non-parameterized, predictions of galaxy angular size, a self-consistent test is still obtained.

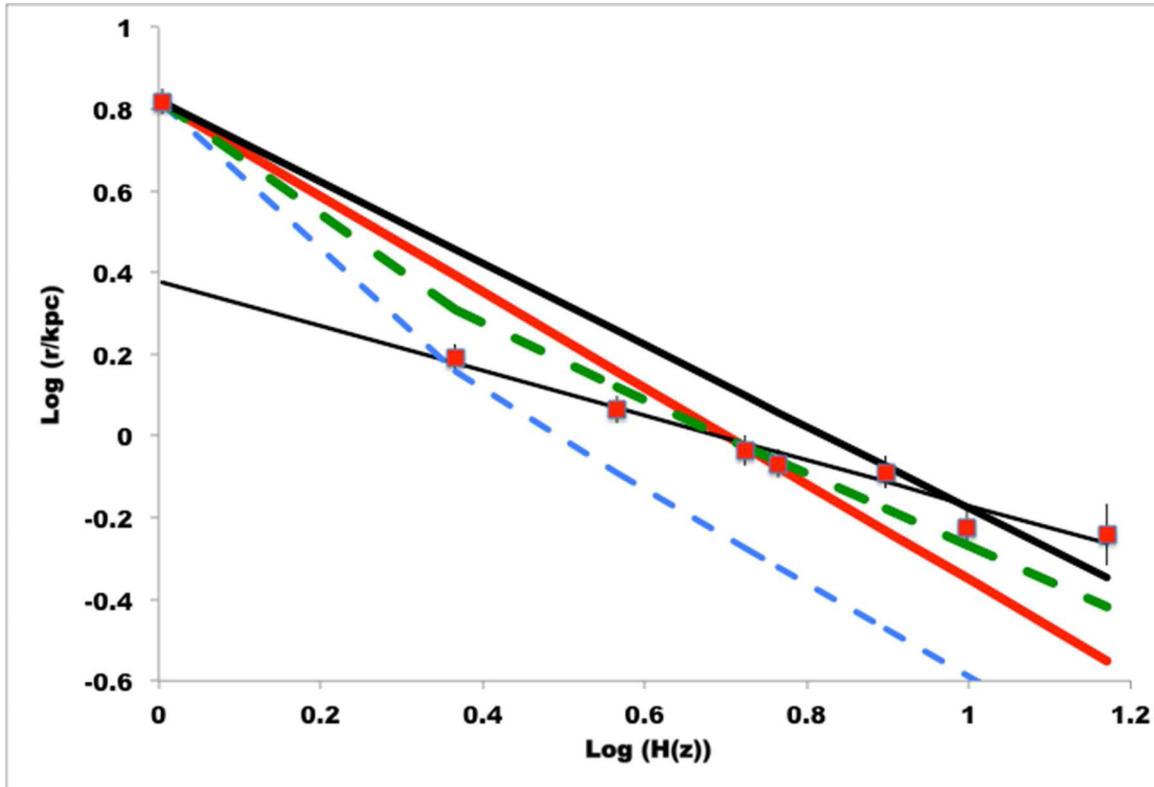

**Figure 2.** The log of the median radii of UV-bright disc galaxies $M \sim -18$ from Shibuya et al. 2015 and the *GALEX* point at $z = 0.027$ from Lerner, Scarpa & Falomo 2014 is plotted against log of $H(z)$, the *Hubble* radius at the given redshift. The expanding universe hypothesis is assumed here, with the concordance LCDM assumptions. The upper heavy black line shows the best fit of the Mo et al. theory of size expansion, which do not fit the data. The lower heavy red line is the best linear fit to the data, with slope of log $r$ on log $(H(z))$ of $-1.13$, which is also a very poor fit to the data. The light black line shows that a good linear fit with slope $-0.55$ is possible only if the low-$z$ point is excluded. The heavy dashed green line shows the fit of $r = (1 + z)^{-1.31}$, still not a good fit and not predicted by any theory. The blue light dashed line $r = (1 + z)^{-1.695}$ illustrates that the two functions $(1 + z)^{-(1.5n)}$ and $H(z)^{-n}$ give similar slope predictions only for $z > 1$.

For the concordance model, the effect is hypothesized to be less significant, as the cosmological magnification of high-$z$ galaxies will make their angular sizes larger relative to telescope resolution. Since by observation, the actual high-$z$ galaxies observed are small in angular radius relative even to *HST* resolution, the effect still must in fact be significant. But it cannot be exactly compensated for, as that would require a unique linear size prediction for the concordance theory. As noted above, the Mo et al. theory provides only a range of predictions. However, we can conclude that the direction of any corrections for the RS effect will make the actual sizes of high-$z$ galaxies smaller than those used in Section 1, and thus would make the observations still further from the Mo et al. predictions.

Since the RS effect is not mentioned in the literature on galaxy size, other than in our own earlier paper (Lerner et al. 2014), it is important to demonstrate clearly that it exists and to examine when it is relevant. We do this here with low-$z$ data from *GALEX* and Sloan Digital Sky Survey (SDSS). In Fig. 3, we present half-light radii as measured in the FUV band for *GALEX* observations of galaxies observed with both SDSS and *GALEX*. Galaxies flagged with FUV artefacts and those with FUV magnitude errors $> 0.2$ were excluded, and we limited the samples in each $z$-range to those with sersic index $< 2.5$ and either $-18 >$ FUV M $> -17$ (small symbols) or $-19 >$ FUV M $> -17.7$ (large symbols). The linear sizes shown are those calculated using the concordance cosmology assumptions. The angular sizes are taken from the SDSS and *GALEX* catalogues (DR7 release, http://skyserver.sdss.org/dr7/en/).

A number of points are evident. For the smaller, less bright galaxy samples, the measured median radius increases by a factor of 2 comparing $z = 0.04$ and $0.14$ for either *GALEX* or SDSS measurements. By contrast with galaxy size scaling as $(1 + z)^{-1.5}$, the expectation would be a *decline* in size by 15 per cent. For the larger, brighter galaxies, the radius increase in the same $z$ range is also significant, but less, a factor of 1.6. For these larger galaxies, it is clear that the RS effect is no longer relevant for distances less than about $z = 0.07$, which for *GALEX* are median angular radii greater than about twice the effective resolution. It is also clear that the effect is, as expected, much more pronounced for both smaller galaxies and for larger resolutions (*GALEX* vs SDSS).

The larger measured sizes can be explained by the resolution effect, with galaxies small compared with the *GALEX* resolution of about 2.5 arcsec being measured too large, an effect that becomes greater as $z$ increase and the angular sizes of the galaxies become smaller. However, the fact that SDSS excludes galaxies smaller than 2 arcsec in the $r$ band also could produce a selection effect by eliminating some of the small-angular size galaxies as measured in the FUV.

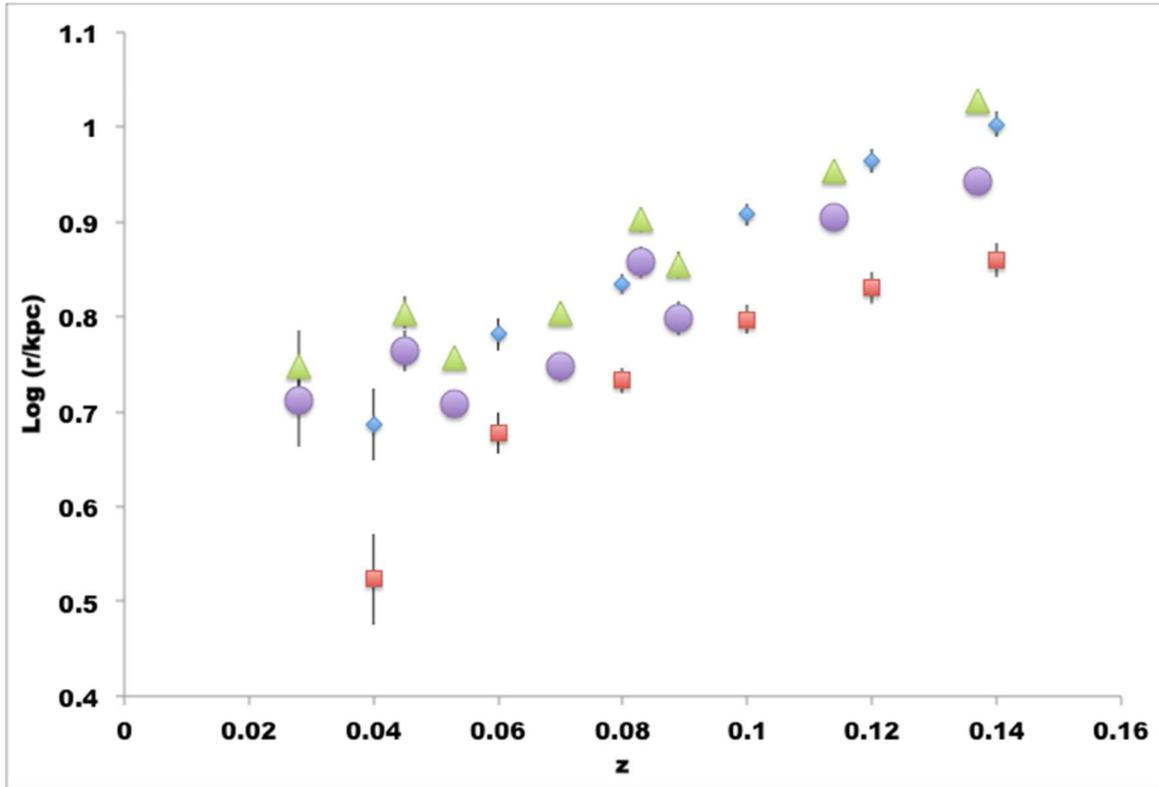

**Figure 3.** Galaxies with sersic index < 2.5 and −18 >FUV M>−17 (small symbols) or −19 > FUV M>−17.7 (large symbols) and show measured radii that increase sharply with distance at low redshift. The same samples of galaxies measured with SDSS (red squares and purple circles) have consistently smaller measured radii than when measured with the lower resolution *GALEX* (green triangles and blue lozenges).

We can separate these two influences by comparing the measurement of the *same* galaxies in the UV with *GALEX* and with the higher resolution SDSS, with an effective resolution of less than 1 arcsec. We limit the sample to galaxies with $z < 0.1$. In Fig. 4, we plot the 20-galaxy running average of log *GALEX* radii for disc galaxies ordered by ascending SDSS radius. As can be seen, the ratio of the two measurements strongly diverges from unity (straight line) for SDSS radii less than about 4 arcsec, or 1.6 times the *GALEX* resolution. It is also notable that the measured *GALEX* radii level off at 3 arcsec, indicating that median radius measurements of samples are probably not very accurate for angular radii smaller than the telescope resolution.

The measurement techniques used in the SDSS and *GALEX* catalogues are not identical nor are the wavelengths measured (300–400 nm for SDSS and 138–165 nm for *GALEX*). However, the strong non-linear dependence of the ratio of the measured radii on the *angular* radius of the galaxy confirms that this effect is due to the different resolutions of the two surveys.

The RS effect can, in fact, be expected, as galaxies that are small in angular extent compared with a telescope resolution will have nearby neighbours blurred into a single profile, increasing apparent radius, while with larger galaxies the neighbours will be resolved and excluded from the profile.

## 4 SEU HYPOTHESIS FOR DISC GALAXIES

We can take into account the RS effect by matching the linear resolutions in a comparison of high-$z$ and low-$z$ galaxies, as we did in previous work (Lerner et al. 2014). As described in detail in that work, we match *GALEX* low-$z$ samples with HUDF high-$z$ samples using identical algorithms to fit exponential light profiles. For the original data used there, each HUDF sample with a redshift of $z$ is matched with one with redshift of $z/38$, so that, in the SEU model, the linear resolution of *GALEX* is the same as HUDF. As shown in that paper, the effective resolution of *HST* is 1/38 that of *GALEX*. It is important to note that we use effective resolution, defined in detail in that paper as the smallest radius for an object that can be distinguished from a point source, so this effective resolution depends on both the telescope and on the method of image analysis used for identifying point sources. The samples and the HUDF filter were selected so that the rest-frame bandwidths were also closely matched, as were the mean absolute luminosities.

In Fig. 5, we compare the difference in median log *r*, (log *r* HUDF − log *r* GALEX) for each of the eight pairs, showing the same data in Table 2. As can be seen, the data are excellent fit to the parameter-free prediction of no change in log *r*.

We also can compare the newer, and more extensive data from Shibuya et al. Since the $z$ of these samples is not exactly matched to the $z$ of the *GALEX* samples, and thus to the resolution of the *GALEX* telescope, we interpolate the *GALEX* median r results by fitting a quadratic function to these results, $r$ (kpc) $= 411z^2 - 6.896z + 5.73$. Again we compare the Shibuya et al. *HST* results at a given $z$ with the *GALEX* interpolation function for $z/38$.

The results are also shown in Fig. 5 and Table 2. In Table 2, the third column is the statistical error in the sample median radius difference between the *GALEX* and HUDF samples. Taking the two data sets together, the predictions and observations are remarkably close. The mean difference in log *r* of all samples is 0.0086. In other words, the mean of the high-$z$ radii and the low-$z$ radii differs by 2 per cent. Taken together, the points are fit by log $r = (0.016 \pm 0.023)$ log $(1 + z)$, again statistically an excellent fit to the prediction of zero slope. The root mean square of the difference in log $r$ of individual comparisons is 0.035 and the chi square for the fit to zero difference is 16.0, a completely acceptable value for 13 degrees of freedom.

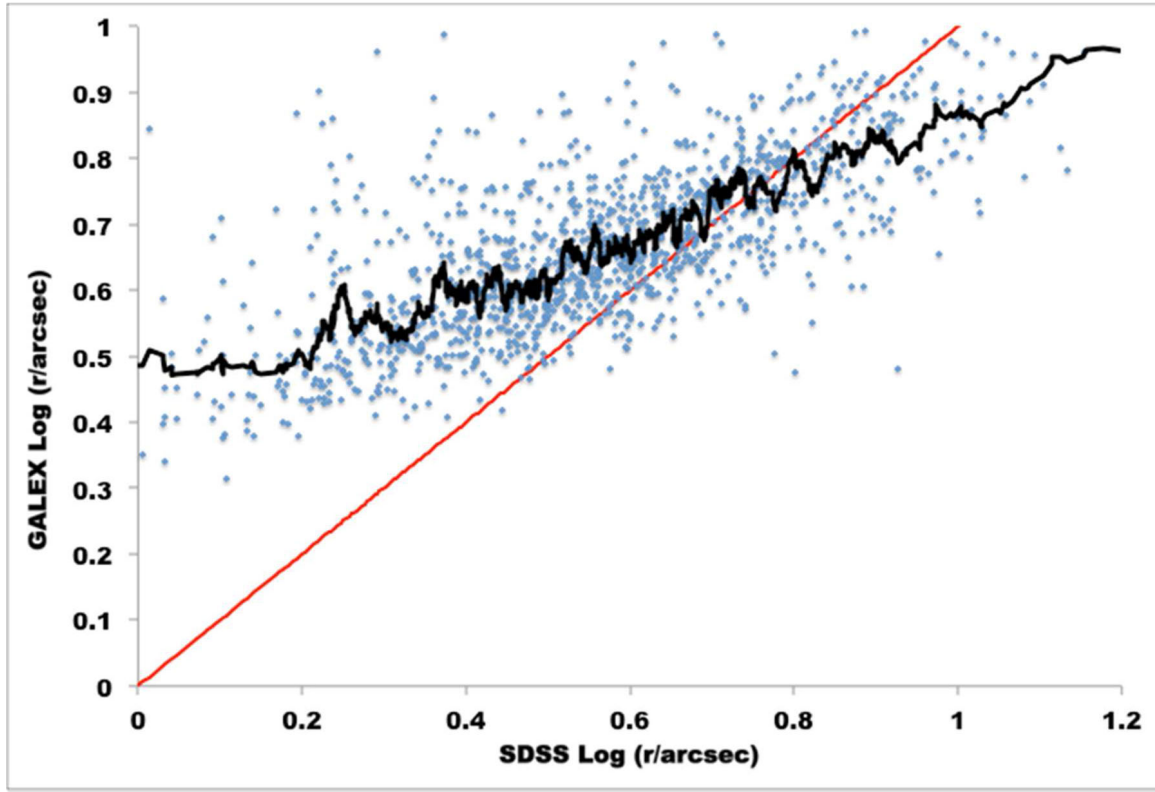

**Figure 4.** The log angular radius as measured in the *GALEX* catalogue is plotted against the log angular radius as measured by SDSS for the same galaxies. Individual galaxies are blue dots, while the black curve is a 20-galaxy running average, ordered by SDSS angular radius. For comparison, the straight red line represents identical radii for both instruments.

Thus, these data are fit by the SEU hypothesis predictions with no free parameters. But predictions based on the expanding universe hypothesis provide no acceptable fit to the same data, despite having an additional free parameter, the slope of log $r$ on log($H(z)$). We emphasize that not only are these data compatible with the SEU hypothesis, but the fit to the zero slope line on the axis has to be viewed as a large coincidence from the standpoint of the expanding universe hypothesis.

It is important to note here, as was described in greater detail in Lerner et al., that in comparing these UV-bright disc galaxies we are comparing galaxies with similar very young stellar populations, so we can safely ignore any evolutionary changes that may occur in individual galaxies.

## 5 SIZE EVOLUTION HYPOTHESIS FOR ELLIPTICAL GALAXIES

For elliptical galaxies, which have evolved stellar populations, we are unable to completely cancel out evolutionary effects in the same way we can for UV-bright disc galaxies. However, by comparing size for both the most massive and the most luminous elliptical galaxies in each redshift bin, we can adequately test for the effect of such evolution. A strong evolutionary effect will create a difference between the size–redshift relationship of the galaxy populations with the greatest stellar mass and those with the highest luminosity. If we observe no such differences, we can again safely ignore evolutionary effects of individual galaxies, which would have no net effect on the mass–radius or luminosity–radius relationships of entire galactic populations. We note that since the method for estimating stellar masses depends linearly on the measurement of absolute luminosity, the predictions for the two cosmological models are the same for the two measurements.

Fortunately, there exists a comprehensive data collection of elliptical galaxy size that is binned both by stellar mass and by luminosity (van der Wel et al. 2014). We select from this source the most massive and the brightest ellipticals that can be plotted over the whole redshift range provided of $z = 0.25$–2.75. This is log $M = 11$ solar masses and $V$-band log $L = 10.75$, derived from Table 1 and Table A2, respectively.

We show these results for the concordance expanding universe hypothesis in Fig. 6. In the figure, we include as well the disc galaxies from Fig. 2. It can be seen that the ellipticals lie on a virtually identical line to the disc galaxies. From the standpoint of the expanding hypothesis, this is again a surprising coincidence since the size evolution of the two types of galaxies is supposed to proceed from entirely different processes. In the case of discs, it is the growing size of the dark matter halo at the time of formation of the galaxies, while for the ellipticals it is the growth in radius of individual galaxies, getting older and larger with decreasing $z$. We can also see that the size–z relationship is broadly similar for the galaxies binned by luminosity (large dots) and by mass (small dots).

As with the disc galaxies, the steepest Mo et al. relationship, log $r = -\log(H(z))$, is incompatible with the data. However, the elliptical luminosity bin data can be fit by log $r = n \log(H(z))$, where $n = 1.44 \pm 0.04$ (chi squared 6.4) or by log $r = n \log(1 + z)$, where $n = 1.55 \pm 0.09$ (chi squared 3.9). Similarly, the data binned by

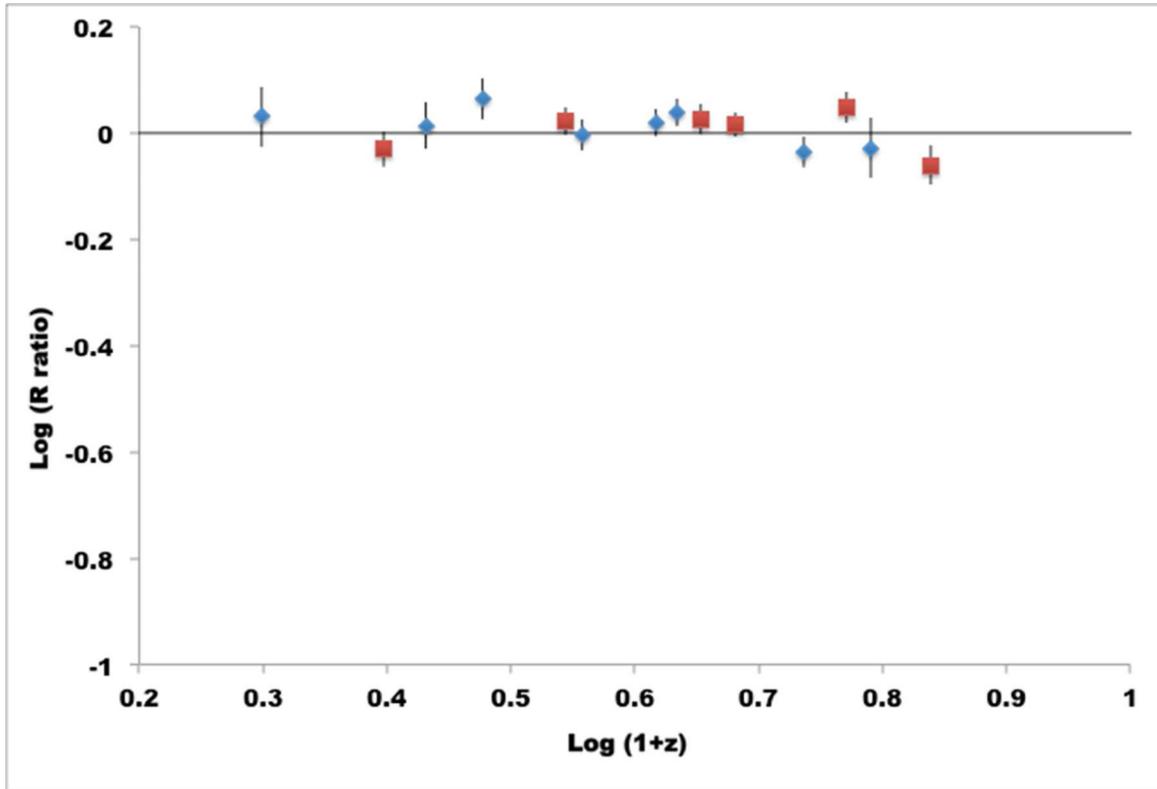

**Figure 5.** The same data as in Fig. 1 are plotted as log of the ratio of mean radius of the high-$z$ sample to the mean radius of the low-$z$ sample against log $(1 + z)$ of the high-$z$ sample, assuming the SEU hypothesis, a non-expanding universe with linear *Hubble* relation. The data an excellent fit to the prediction of no size evolution.

**Table 2.** Difference in median radii of bright disk galaxies HUDF-GALEX (non-expanding assumptions).

| log(1 + z) | Difference in log R HUDF–GALEX | Error | Source |
|---|---|---|---|
| 0.299 | 0.031 | 0.056 | Lerner et al. (2014) |
| 0.431 | 0.014 | 0.043 | Lerner et al. (2014) |
| 0.477 | 0.064 | 0.038 | Lerner et al. (2014) |
| 0.558 | –0.003 | 0.030 | Lerner et al. (2014) |
| 0.617 | 0.019 | 0.025 | Lerner et al. (2014) |
| 0.634 | 0.039 | 0.026 | Lerner et al. (2014) |
| 0.736 | –0.036 | 0.028 | Lerner et al. (2014) |
| 0.791 | –0.028 | 0.056 | Lerner et al. (2014) |
| 0.398 | –0.03 | 0.034 | Shibuya et al. (2015) |
| 0.544 | 0.02 | 0.027 | Shibuya et al. (2015) |
| 0.653 | 0.026 | 0.028 | Shibuya et al. (2015) |
| 0.681 | 0.015 | 0.023 | Shibuya et al. (2015) |
| 0.771 | 0.048 | 0.029 | Shibuya et al. (2015) |
| 0.839 | –0.06 | 0.038 | Shibuya et al. (2015) |

*Notes.* Same data as in Fig. 5. Column (1): mean log (1+z) of HUDF sample. Column (2): Difference in median log radius of galaxies, HUDF-GALEX samples. Column (3): Statistical error in Column (2). Column (4): source of HUDF sample.

stellar mass can be fit by $\log r = n \log(H(z))$, where $n = 1.34 \pm 0.04$ (chi squared 7.1) or by $\log r = n \log(1 + z)$, where $n = 1.47 \pm 0.09$ (chi squared 3.0). We again emphasize that there is no theoretical prediction either of $n$ as large as 1.34 for the dependence on log $(H(z))$ nor for any linear dependence at all on $\log(1 + z)$.

So we can conclude that, assuming the concordance cosmology, the luminosity-binned and mass-binned data follow approximately the same relationship, but both contradict the Mo et al. predictions, although they can, unlike the disc galaxies, be fit by $\log r = \sim 1.4 \log(H(z))$.

## 6 SEU HYPOTHESIS FOR ELLIPTICAL GALAXIES

If we test the SEU hypothesis against the elliptical disc data, we do not have to take into account the RS effect. As shown in Section 3 above, the RS effect is not significant for median angular radii greater than about twice telescope resolution. Because the elliptical data do not extend beyond $z = 2.75$, the median angular radius for all samples is more than 0.13 arcsec, twice the effective *HST* resolution of 0.066 arcsec. So the RS effect can be ignored.

In Fig. 7, we plot the log ratio of the median SEU radius for the sample at the given $z$ to the median SEU radius for the lowest $z$ samples at $z = 0.25$. The same data are given in Table 3. Again, the large black dots are for the luminosity-binned samples and the small black dots for the stellar-mass binned samples. As can be seen, once again the data are excellent fit to the SEU prediction of no change in radius. The mean difference in log $r$ of all samples is 0.007. Taken together, the points are fit by $\log r = (0.042 \pm 0.08) \log(1 + z)$, again statistically an excellent fit to the prediction of zero slope. The root mean square of the difference in log $r$ of individual comparisons is 0.04 and the chi square for the fit to zero difference is 9.1, an excellent value for 9 degrees of freedom. There is no statistically significant difference between the results for the luminosity-binned and mass-binned samples taken separately. For example, the best-fitting slope on log(1 + $z$) is 0.037 + 0.085 for the luminosity binned samples and 0.06 ± 0.1 for the mass-binned samples. Thus, again, with no free parameters, the SEU prediction fits the elliptical galaxy size data. Given the correspondence of the mass and luminosity binned results, no evolutionary effect needs to be considered.

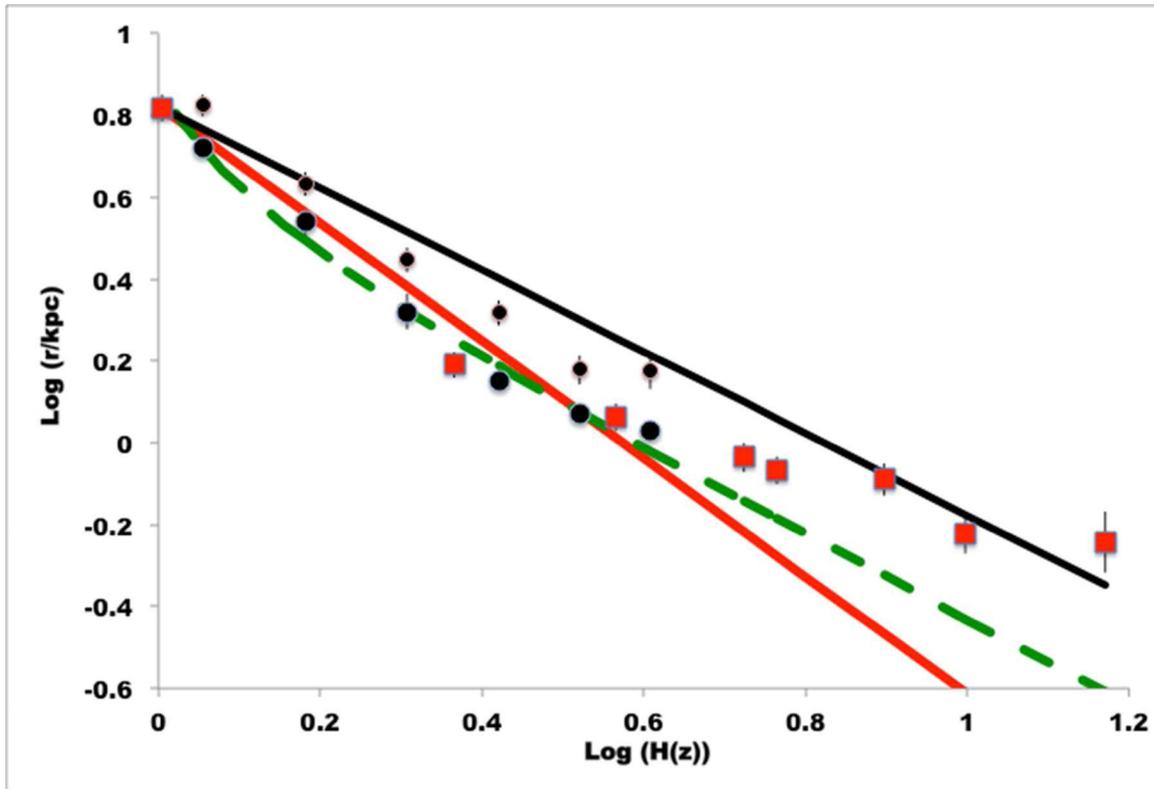

**Figure 6.** Elliptical galaxies from van de Wel et al. 2014 are plotted with log radius in the concordance cosmology against log($H(z)$). Large dots are galaxies with mean log $L$ = 10.75 while small dots are galaxies with mean log $M$ = 11, where $L$ is $V$-band luminosity and $M$ is stellar mass, both in solar units. As in Fig. 2, red squares are disc galaxies. Note that elliptical galaxies lie along the same curve as disc galaxies and thus lie far below the steepest dependence on log($H(z)$) predicted by the Mo et al. theory (black upper line). Best-fitting curves are shown for log $r = n$log($H(z)$) with $n$ = 1.44 (lower red line) and log $r = n$log(1 + $z$) with $n$ = 1.55 (dashed green curve).

In Fig. 8, we combine the elliptical and disc galaxy luminosity-binned data. As can be seen, the whole data set is an excellent fit to the SEU hypothesis. The chi-squared for the combined sample is 21 for 20 degrees of freedom. The mean difference in log $r$ of all samples is 0.0041. The best-fitting slope on log(1 + $z$) is 0.016 ± 0.033, again indistinguishable from the predicted zero slope.

## 7 DO ELLIPTICAL AND DISC GALAXIES HAVE THE SAME R–Z RELATIONSHIP?

This analysis makes clear that the size–redshift relationship of elliptical and disc galaxies can be described by a single simple relationship of no change in the SEU cosmology. But we can also show, using the concordance assumptions, the opposite, incorrect, conclusion can be arrived at – that disc galaxies and elliptical have *different* size–redshift relationships, as other authors (Trujillo et al. 2007; Buitrago, Trujillo & Conselice 2010) have concluded.

If observations were not affected by the RS effect, a zero change in radius with SEU assumptions would imply a single $r$–$z$ relationship with concordance assumptions. In this case, the relationship would be very close to log $r$ = −1.5 log (1 + $z$), where $r$ is the ratio of a sample median radius at a redshift of $z$ to the median radius of a matched sample at $z$ = 0. This relationship follows from the fact that, for all expanding universe models, SB decreases exactly as $(1 + z)^3$ and that, for concordance assumptions, the absolute luminosity in AB units for a source with a given apparent luminosity scales almost exactly as $z^2$, that is as it does with SEU assumptions. As pointed out in Section 1, the actual difference between the two luminosity scales would cause a deviation from the −1.5 log(1 + $z$) relationship of less than 0.04 in log $r$.

However, in the real situation, as detailed in Section 3, observations can be heavily affected by the RS effect if the median angular radius of the sample is less than about twice the resolution of the telescope used. This effect means that smaller angular radius samples, generally those at great $z$, are systemically observed to be larger than the same galaxies observed with finer resolution. The net result, if the RS effect is not taken into account, is to flatten the $r$–$z$ slope of galaxy samples at higher $z$, and of galaxy samples with smaller linear radii.

The consequence for comparing ellipticals and discs is clear by comparing Figs 2 and 6. Although the disc galaxies and elliptical clearly lie along the same curve, the ellipticals are observed over a smaller redshift range. Thus, over the shorter range, the elliptical can be well fit by a steep slope on log(1 + $z$), very close to –1.5. So can the discs in this red-shift range. However, over the larger redshift range, the discs cannot be well fit by either function, unless low-$z$

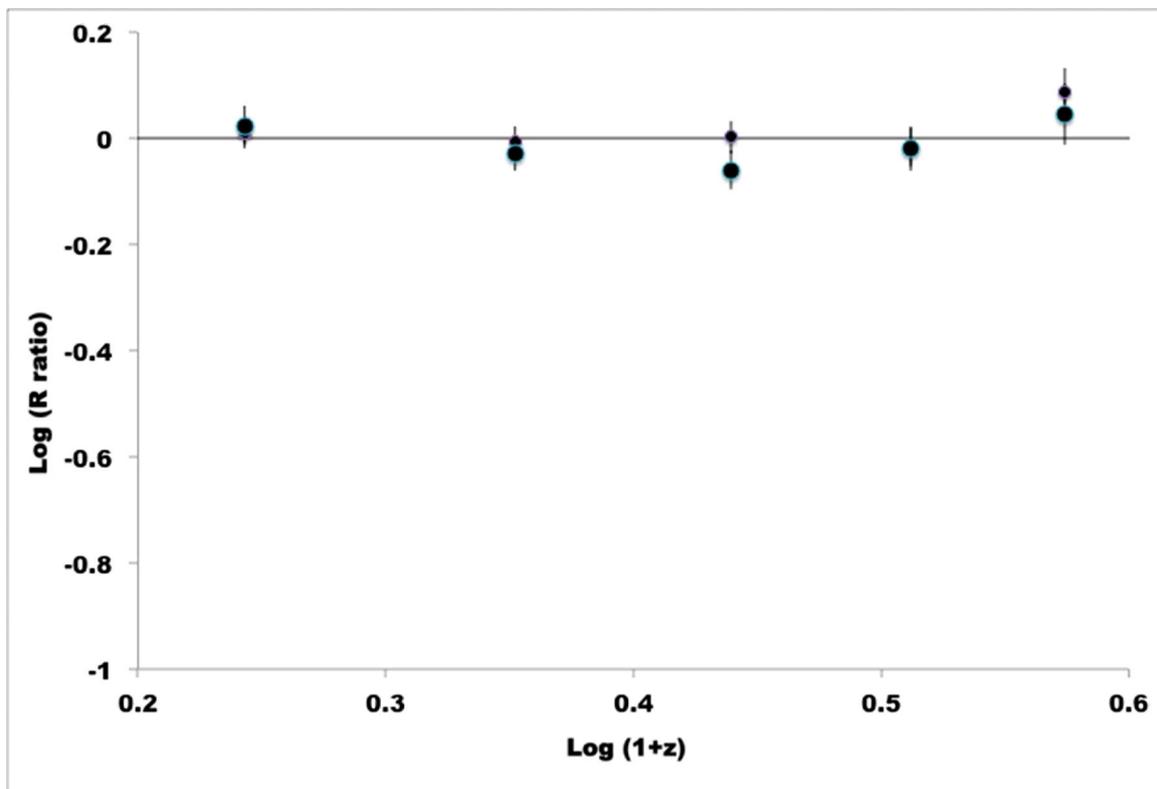

**Figure 7.** The elliptical data in Fig. 6 are plotted assuming SEU hypothesis as log of the ratio of mean radius of the high-$z$ sample to the mean radius of the $z = 0.25$ sample against log $(1 + z)$ of the high-$z$ sample. Large dots are for samples binned by luminosity, small dots for those binned by mass. Where only one dot is visible at a given log$(1 + z)$ the difference between the two samples is smaller than the size of the symbol.

**Table 3.** Difference in median radii of elliptical galaxies high-$z$ vs low-$z$ samples (non-expanding assumptions).

| log$(1 + z)$ | Mass bin log $R$ diff | Error | Luminosity bin log $R$ diff | Error |
|---|---|---|---|---|
| 0.243 | 0.011 | 0.030 | 0.024 | 0.036 |
| 0.352 | −0.008 | 0.031 | −0.030 | 0.032 |
| 0.439 | 0.002 | 0.031 | −0.060 | 0.036 |
| 0.512 | −0.018 | 0.036 | −0.020 | 0.042 |
| 0.574 | 0.087 | 0.045 | 0.046 | 0.058 |

*Notes.* Same data as in Fig. 7. Column (1): mean log$(1+z)$ of high-$z$ sample. Column (2): Difference in median log radius of galaxies, high $z$ sample vs sample at $z = 0.25$ for galaxies with log $M$ =11. Column (3): statistical error in Column (2). Column (4): Difference in median log radius of galaxies, high $z$ sample vs sample at $z = 0.25$ for galaxies with $V$-band log $L$ = 10.75. Column (5): statistical error in Column (4).

measurements are incorrectly omitted. A best fit to all the spiral data thus arrives at a smaller slope, although with an unacceptable fit. The impossibility of a good fit to the $z > 3$ data also accounts for the wide range of slopes claimed for discs galaxies, as compared to the narrow range for ellipticals. As shown in Section 4, the higher $z$ measurements are shifted to larger values by the RS effect. If the effect is ignored, a spurious difference in slopes results, while if it is correctly taken into account, as we do in Section 4, all galaxies can be fit by the line of zero change in SEU, which is mathematically equivalent to a single fit of −1.5log $(1 + z)$ for the concordance hypothesis.

In some cases, such as van der Wel et al., different slopes for discs and ellipticals are reported even for $z < 3$. Here, the reported sizes for discs clearly are in conflict with that for other sources, such as Shibuya et al. Thus, for example, van der Wel et al. show the largest bin late-type galaxies at $z = 2.5$ at 3.9 kpc, Shibuya et al. shows the largest disc galaxies as 2.35 kpc, a difference of 0.22 in log $r$, enough to account for a difference of 0.4 in the slope of log $r$ on log $(1 + z)$. As shown in Section 2, the observations reported by Lerner et al. closely agree with those of Shibuya and therefore also disagree with those of van der Wel et al.

This discrepancy can again be accounted for by the RS effect. van der Wel et al. use measurements at constant observed wavelength bands, 1250 and 1600 nm, which at $z = 3$ are about twice the wavelength bands as that used by the two other papers. A complex k-correction is then used to adjust the calculated size. Since disc galaxies, but *not* ellipticals, are observed to be considerably smaller at longer wavelengths, the RS effect is important at smaller redshifts for longer wavelengths, so has distorted the calculated sizes. We can estimate the magnitude of the effect by using the size–wavelength relations from van der Wel et al. (eq 1) for $z = 0$, which implies that doubling the observed wavelength for late-type galaxies leads to a decrease in observed radius of a factor of about 1.5. With the SEU assumption of constant radius, this would mean a median angular radius of only 0.086 arcsec, only 1.3 times *HST*'s effective resolution. From the *GALEX* data, we can see that the RS effect would shift the observed radius of such galaxies up by a factor of about 1.5 or a difference in log $r$ of 0.18, very nearly accounting for the discrepancy in observed values between Shibuya et al. and van der Wel et al. This phenomenon does not affect the elliptical measurements in van der Wel et al. because elliptical galaxies have very close to the same radii as observed at a wide range of wavelengths.

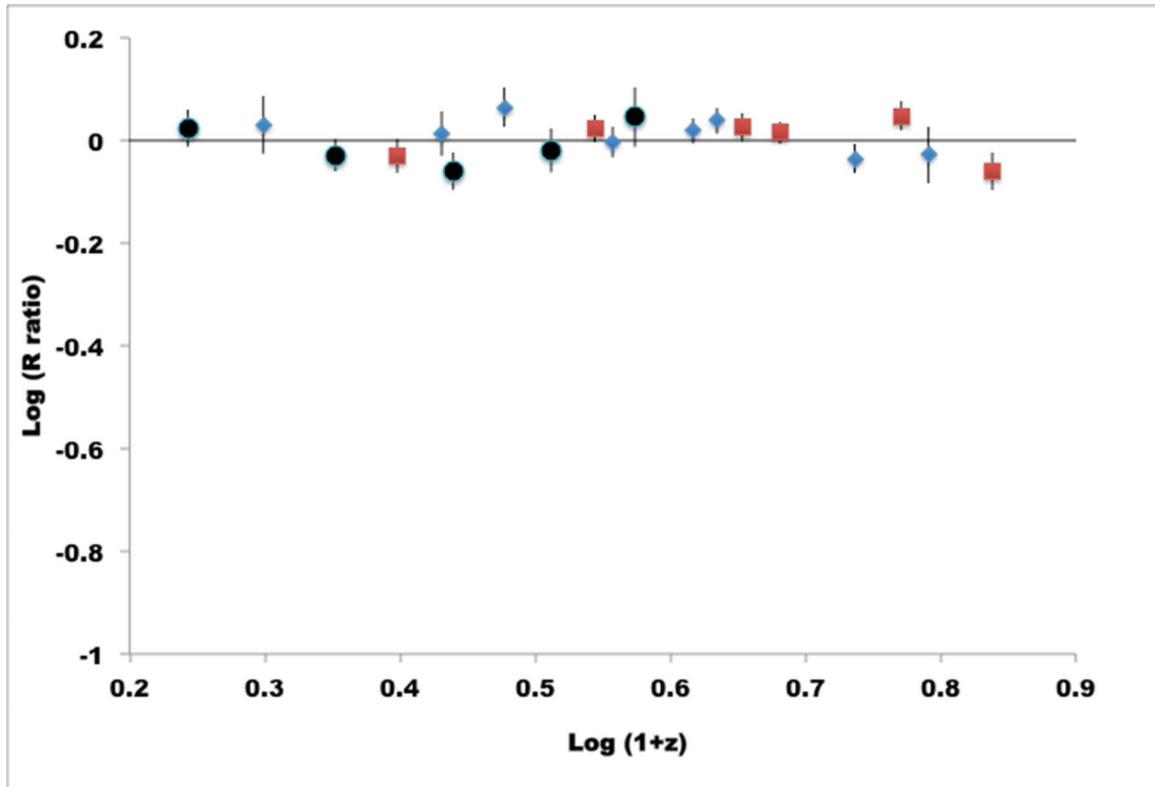

**Figure 8.** Luminosity-binned disc and elliptical galaxy data from Figs 5 and 7 plotted together, again an excellent fit to SEU prediction of no change in radius for given luminosity, which is the same as no change in SB.

Finally, the RS effect of course has a greater impact on samples of galaxies that are intrinsically smaller. If the effect is not taken into account, this result in a spurious reduction of the slope of the log $r$–log$(1 + z)$ relationship for smaller galaxies, which also decreases the slope for disc galaxies taken as a whole. Thus, for example, in van der Wel et al. late-type galaxies with mean log mass = 10.25 have a slope of log $r$ on log$(1 + z)$ of only –0.32, even smaller than the reported slope of –0.75 for the whole late-type sample.

In contrast, if the samples with the largest galaxies are used and the RS effect is correctly taken into account using the SEU hypothesis, the single relationship of no change is valid for both galaxy types, as shown in Fig. 8. As we pointed out above, the RS effect can be used with concordance assumptions only if a specific predicted $r$–$z$ relationship is hypothesized. The resolutions of low-$z$ and high-$z$ telescopes can then be matched for the assumed relationship, which can then be consistently tested. Mathematically, from the above analysis, it follows that only if a relationship very close to log $r = -1.5$ log$(1 + z)$ is assumed with expanding universe assumptions will there be correspondence of predictions with observations once the RS effect is taken into account. In that case, it follows that, as in the SEU case, the relationship will fit both elliptical and disc galaxies.

It is worth emphasizing that, in the SEU assumption, the prediction of constant SB (or constant radius for a given luminosity) comes directly from the non-expanding universe hypothesis. A relationship of log $r = -1.5$ log$(1 + z)$ in the expanding universe hypothesis is mathematically identical (except for the small corrections mentioned above) to constant $r$ for the SEU hypothesis. However, that relationship is not a prediction of *any* physical process that is hypothesized in the literature.

Thus, we can conclude, purely on the basis of the size data, that the predictions that have been made based on the expanding universe hypothesis are not validated. No fit of the form log $r = n$ log $(H(z))$ with $1 > n > 2/3$ is valid for either disc or elliptical galaxies. Disc galaxies cannot be fit with the predicted formula no matter what $n$ is and elliptical galaxies can be well fit only with an $n$ of 1.34, well outside the predicted range, and for which no physical justification was provided in the literature.

## 8 DOES ANY PHYSICAL MECHANISM CORRECTLY PREDICT SIZE EVOLUTION?

We have shown that the Mo et al. predictions are not confirmed for either disc or elliptical galaxies. There appears to be for disc galaxies no other theory of size evolution. However, for the size-evolution hypothesis for ellipticals, a number of physical mechanisms have been examined. The 'puffing-up' hypothesis (Fan et al. 2008) posits a loss of mass due to galactic winds created by quasars. This theory suffers from at least two fatal flaws. First, the amount of gas expelled needed to lead to the observed expansion is far beyond any observations of gas mass in elliptical galaxies, which are gas poor. Assuming angular momentum does not change, the radius of the galaxy will increase as $M^{-3}$. Since the data show a change in radius by a factor of about 8, this implies an initial gas mass fraction of about 0.5. There is no evidence for gas fractions at all close to this value. Secondly, as other authors have pointed out, if quiescent galaxies, undisturbed by major mergers, are steadily emitting gas mass, there should be a correlation between the age of stellar populations and radius for a given stellar mass. But no such relationship exists (Trujillo et al. 2009).

The 'major merger' hypothesis[9] proposes that elliptical galaxies expand as angular momentum is added due to major mergers with galaxies of comparable size. As Taylor et al. (Taylor et al. 2010) point out, there is a fatal flaw with this hypothesis as well. Since only about 1/5000 of existing elliptical galaxies have radii at all comparable to those hypothesized (given an expanding universe) for those at $z = 2.3$ something of the order or 20 collisions will be needed to reduce the undisturbed population by that amount. But observations, for example, Man et al. (Man, Zirm & Toft 2014), indicate that the major merger rate since $z = 2.3$ is about an order of magnitude less than that, or about 2 mergers, one major and one minor on average. This is confirmed by Lopez-Sanjuan et al. (Lopez-Sanjuan et al. 2014), who find 0.57 mergers for quiescent and 0.26 mergers for star-forming galaxies since $z = 1$. An even lower rate of 0.5 mergers since $z = 3.5$ is reported by Mundy et al. (Mundy et al. 2017).

This leaves the 'minor merger' theory (Naab, Johansson & Ostriker 2009), which assumes that the angular momentum needed to increase the radii of elliptical galaxies is supplied by mergers with galaxies that are much smaller than the ones actually observed. Various authors have elaborated this hypothesis to make concrete quantitative predictions, Trujillo et al. (Trujillo, Ferreras & de la Rosa 2011), for example, have calculated the number of minor mergers (with mass ratios around 10) needed to duplicate the observed increase in radius of massive ellipticals. Interestingly, they arrive at a figure of about 10 mergers since $z = 10$ and, projecting their formula (17), about 20 since $z = 2.3$, about the same number of mergers needed by the statistical argument of Taylor et al. (Taylor et al. 2010). But again, the number of minor mergers observed from Man et al. and Lopez-Sanjuan et al. is an order of magnitude less than that required. So all three theories suffer from severe quantitative contradictions with observations.

In addition, both major and minor merger theories imply an increase over time of stellar mass in passive, elliptical galaxies. But actual observations, most recently by Bundy et al. (Bundy et al. 2017), show that no such increase in stellar mass has occurred.

## 9 VELOCITY DISPERSION TEST FOR SIZE EVOLUTION OF ELLIPTICAL GALAXIES

There is an independent test for the size evolution hypotheses, which is their prediction for the evolution of velocity dispersions of galaxies of a given stellar mass. Stellar masses can, of course, be measured independently of radius based on spectral energy distribution and total luminosity. In the case of the merger hypotheses, there is no reason to expect that the ratio of dynamical mass (total mass) to stellar mass should evolve greatly with $z$. For a given stellar mass, then, the velocity dispersion is predicted by these hypotheses to vary as $r^{-0.5}$, and thus to increase with $z$ very significantly, by about a factor of 2.7 by $z = 2.3$. In the case of the 'puffing-up' hypothesis, the ratio of total mass to stellar mass is expected to increase with $z$, and therefore velocity dispersion will increase faster than $r^{-0.5}$. By contrast, for the SEU, non-expanding hypothesis, where the radius of elliptical galaxies of a given stellar mass is expected to be constant, the velocity dispersion will also be constant.

It is possible to compare these predictions and observations. There is a range in observations of evolution of velocity dispersion with redshift. Some studies (Fernández Lorenzo et al. 2011) actually show a small decline in velocity dispersion with increasing redshift. However, to be most generous with the expanding universe hypothesis, we select the study (Cenarro & Trujillo 2009) with the maximum increase in velocity dispersion with $z$. This amounts to a factor of 1.33 at $z = 1.6$. This would be expected for a decrease in radius of a factor of 1.76. But in actuality, if we assume the expanding universe hypothesis, the observed decrease in radius is actually a factor of 3.5. In other words, if we assume the expansion hypothesis, this requires that the dynamical mass at this redshift be a factor of 2 less relative to stellar mass than for the present-day universe. Other studies, such as Di Teodoro, Fraternali and Miller (Di Teodoro, Fraternali & Miller 2016), find no change at all in velocity dispersion in disc galaxies at $z = 1$, as predicted by SEU assumptions.

Some authors (Trujillo, Ferreras & de la Rosa 2011) have hypothesized that the ratio of dark matter to baryonic matter is increasing with time in elliptical galaxies, although theories published prior to these observations deduced the opposite, that galaxies started out as purely dark matter concentration that baryonic matter fell into. However, these additional hypotheses are contradicted by the fact that, in higher $z$ elliptical galaxies, the ratio of dynamical mass to stellar mass fall well below unity and continues to decline with increasing $z$. Peralta de Arriba et al. (Peralta de Arriba et al. 2015) report that, using stacked spectra to determine velocity dispersions, galaxy samples at $z > 0.5$ have $M_{dyn}/M_{stell}$ as low as 0.4. Clearly this is physically impossible – the total mass cannot fall below the stellar mass (unless one wants to hypothesize a new form of dark matter with negative mass!). These authors do argue that an increase in the virial coefficient could account for the data, but the increase, which would have to be by a factor of 5 by $z = 2$, cannot be accounted for by any physically possible model.

Therefore, as with disc galaxies, we are forced to conclude that the predictions based on the size-evolution, expanding universe hypothesis are incompatible with the observations for elliptical galaxies. All three mechanisms for size evolution are contradicted by observations, and *any* size evolution hypothesis leads to the physically impossible situation that total mass of the galaxies is less than stellar mass.

Since the radius does not change for a given stellar mass, the SEU hypothesis predicts a constant velocity dispersion for a given stellar mass, irrespective of $z$. As noted above, there is a range of measurements for the dependence of velocity dispersion on $z$, but at the present time, constant velocity dispersion falls with that range. More significantly, the SEU hypothesis completely resolves the unphysical excess of stellar mass over dynamical mass. Using the SEU values for radius, and thus for dynamical mass, all the stacked galaxy values of the ratio of dynamical mass to stellar mass in Peralta de Arriba are $>1$, as is required to be physically possible.

## 10 DISCUSSION—THE KEY ROLE OF PREDICTIONS

In this paper, I have shown that the predictions that have been made on the basis of the expanding universe hypothesis are incompatible with the data for galaxy size for a given luminosity, or equivalently for SB. The emphasis that I have placed here on comparing actual predictions with data would in most scientific contexts require no explanation. It is a universally accepted truism that the test of the scientific validity of a hypothesis is whether predictions based on the hypothesis are in accord with, or contradicted by, actual observations made after the predictions. However, in cosmology it has unfortunately been the case that even a long series of failed predictions has not generally led to the rejection of theories, but rather to their unlimited modification with *ad hoc* hypotheses, such as inflation, non-baryonic matter, and dark energy.

The entire history of scientific-technological advance has shown the value of judging theories by their predictions. A hypothetical aircraft, for example, using a theory of aerodynamics that needed to be fine-tuned after every few miles of flight would scarcely be useful. Even in cosmology, apparently far removed from technological applications, multibillion-dollar public investments in astronomical instruments are in fact based on the predictions of cosmological theory. In cosmology, as in the rest of science, judging theories on the basis of their predictions is the only real way towards scientific knowledge.

Of course, broad hypotheses such as that of an expanding universe, the much more detailed concordance cosmology model, or an alternative such as the SEU hypothesis need to have their predictions tested against all available sets of data: for example, those on abundance of light elements, large-scale structure, and the cosmic background radiation. This paper covers only one such set of data – that for galaxy size or, equivalently, SB. Other data sets have also been considered by me and by others for both these and other models (Lerner 1993, 2003; Arp & van Flandern 1992; López-Corredoira 2017). The ultimate acceptance or rejection of such broad hypotheses rests on the compatibility of their predictions with all of these data sets.

## 11 CONCLUSIONS

Predictions based on the size-evolution, expanding-universe hypothesis are incompatible with galaxy size data for both disc and elliptical galaxies. For discs, the quantitative predictions of the Mo et al. theory are incompatible at a $5\sigma$ level with size data, as is any model predicting a power-law relationship between $H(z)$ and galaxy radius. For ellipticals, a power law of $H(z)$ does fit the data, but only with an exponent much higher than that justified by the Mo et al. theory. All three mechanisms proposed in the literature – 'puffing up', major and minor mergers – make predictions that are contradicted by the data, requiring either gas fractions or merger rates that are an order of magnitude greater than observations. In addition, any size evolution model for ellipticals leads to dynamical masses that, given the observed velocity dispersions, are smaller than stellar masses, a physical impossibility.

Contrary to some other analysis, we find that the $r$–$z$ relationships for elliptical and disc galaxies are identical. The RS effect must be taken into account for valid conclusions, and that effect is larger for disc galaxies that have smaller angular radii, either because they are observed at higher $z$ or because they are observed at longer rest-frame wavelengths. The identical size evolution of discs and ellipticals appears as a very large and unexplained coincidence in the expanding-universe model.

In contrast, the SEU model with a linear distance–$z$ relationship is in excellent agreement with both disc and spiral size data, predicting accurately no change in radius with $z$. The exact agreement of the SEU predictions with data could also only be viewed as an implausibly unlikely coincidence from the viewpoint of the expanding universe hypothesis. The contradictions with impossibly small dynamic masses are also eliminated with the non-expanding universe model.

These conclusions of course refer only to the size-redshift data sets. It is important to note that any overall comparisons of cosmological models must be based on all available data-sets.

## ACKNOWLEDGEMENTS

The author would like to thank Dr. Takatoshi Shibuya for his kind help in providing clarifications and additional data and Dr. Martin López-Corredoira for his helpful suggestions.

This paper has been typeset from a Microsoft Word file prepared by the author.